\let\orilabel\label
\documentclass[aps,pra,preprint,a4paper,showpacs,showkeys,superscriptaddress,nofootinbib]{revtex4-1}
\let\label\orilabel

\usepackage{latexsym}
\usepackage{amsmath,amssymb,mathrsfs}
\usepackage{mathtools}
\usepackage{graphicx}
\usepackage{subfigure}
\usepackage{xcolor}
\usepackage{physics}
\usepackage{cancel}
\usepackage{tikz}
\usepackage{multirow,tabularx}

\newcolumntype{C}[1]{>{\centering\arraybackslash}p{#1}}
\usepackage{hyperref}     
\hypersetup{colorlinks,%
  citecolor=blue,%
  linkcolor=cyan,%
}
\usepackage[titletoc]{appendix}
\usepackage{enumerate}
\graphicspath{{./Figures/}}
\usetikzlibrary{decorations.pathmorphing,decorations.pathreplacing,decorations.shapes}
\begin{document}


\title{Validity of black hole complementarity in an accelerating Schwarzschild black hole}

\author{Wontae Kim}%
\email[]{wtkim@sogang.ac.kr}%
\affiliation{Department of Physics, Sogang University, Seoul, 04107,
	Republic of Korea}%
\affiliation{Center for Quantum Spacetime, Sogang University, Seoul 04107, Republic of Korea}%

\author{Mungon Nam}%
\email[]{clrchr0909@sogang.ac.kr}%
\affiliation{Department of Physics, Sogang University, Seoul, 04107,
	Republic of Korea}%
\affiliation{Center for Quantum Spacetime, Sogang University, Seoul 04107, Republic of Korea}%
\date{\today}

\begin{abstract}
Black hole complementarity has been well understood in spherically symmetric black holes.
To study its validity for an accelerating Schwarzschild black hole, which has a preferred direction, we perform the thought experiment proposed by Susskind and Thorlacius and further investigate the criteria set by Hayden and Preskill.
First, we obtain the conserved energy of the accelerating Schwarzschild black hole from the quasilocal formalism, connecting the ADT current with the linearized Noether current at the off-shell level. Using the conserved energy,
we conduct thought experiments based on the Page time and the scrambling time, which show that black hole complementarity remains valid, although the energy required for the duplication of information depends on the angle due to the axisymmetric metric.
\end{abstract}
%



\maketitle


\raggedbottom

\section{Introduction}
\label{sec:introduction}

Black holes have been shown to emit thermal radiation and semiclassically undergo complete evaporation~\cite{Hawking:1974rv,Hawking:1975vcx}. This discovery raises a fundamental question~\cite{Hawking:1976ra}: the loss of quantum information. Several theories have been proposed to address this information loss paradox. These include the idea that the majority of the information is carried away by Hawking radiation ~\cite{Page:1979tc,tHooft:1990fkf,Susskind:1993if,Susskind:1993ki,Susskind:1993mu,Page:1993wv}, the possibility that black holes may not evaporate completely ~\cite{Banks:1992ba,Banks:1992is,Banks:1992mi,Giddings:1992ff}, the suggestion that  significant portion of the information is released during the final stages of black hole evaporation~\cite{Hotta:2015yla}, and the proposal of alternative horizon-free, non-singular structures such as fuzzballs~\cite{Mathur:2005zp,Skenderis:2008qn}.

To resolve the information loss paradox, the idea of black hole complementarity (BHC) was introduced as an attempt to reconcile quantum mechanics with the equivalence principle of general relativity~\cite{Susskind:1993if}. BHC indicates that different observers can have complementary descriptions of black hole dynamics, yet mutually consistent. For instance, a freely falling observer (often referred to as Alice) who crosses the event horizon perceives nothing special at the horizon. In contrast, a stationary observer (commonly called Bob) would perceive something special near the ``stretched horizon'' of the black hole.
Importantly, BHC asserts that these different perspectives do not violate the principles of quantum mechanics.
In Ref.~\cite{Susskind:1993if}, the authors demonstrated that if Bob retrieves Alice's information from Hawking radiation and then falls into the black hole, he would not be able to observe the same information inside the black hole before hitting the curvature singularity.
This is based on the assumption that the information absorbed by the black hole can be retrieved by Bob after the Page time~\cite{Page:1993wv}.
Furthermore, Hayden and Preskill~\cite{Hayden:2007cs} extended this framework by suggesting that if quantum information is absorbed by a black hole after the Page time, it could still be recovered after the ``scrambling time,'' which is shorter than the Page time.
Alice and Bob are located in causally disconnected regions inside the black hole, avoiding any violation of the no-cloning theorem by ensuring that the same information is never accessible to both observers simultaneously~\cite{Susskind:1993if,Hayden:2007cs}.
Some aspects of BHC for various black holes have been explored ~\cite{Kim:2013fv,Chen:2014bva,Gim:2015zra,Gim:2017nnl,Gim:2017rmn,Wu:2022uzy,Kim:2023qvb}; however, all of them pertain to non-accelerating black holes.

For accelerating black holes described by the C-metric~\cite{Kinnersley:1970zw,Plebanski:1976gy,Dias:2002mi,Griffiths:2005qp,Griffiths:2009dfa},
many intriguing structures have been investigated in Kerr-Newman black holes~\cite{Dutta:2005iy,Astorino:2016xiy,Ball:2021xwt,Siahaan:2024ilq}, anti-de Sitter black holes~\cite{Appels:2016uha,Appels:2017xoe,Gregory:2017ogk,Anabalon:2018ydc,Anabalon:2018qfv,Cassani:2021dwa,Kim:2023ncn,Kim:2024dbj,Liu:2024fvq}, NUT black holes~\cite{Podolsky:2020xkf,Podolsky:2021zwr,Podolsky:2022xxd,Barrientos:2023tqb,Siahaan:2018qcw,Astorino:2023elf,Astorino:2023ifg,Siahaan:2024ljt}, and three-dimensional black holes~\cite{Astorino:2011mw,Arenas-Henriquez:2022www,Arenas-Henriquez:2023hur}.
A notable characteristic of these black holes is the existence of at least one unavoidable conical deficit angle along the azimuthal axis. The acceleration of these black holes is driven by conical singularities, which can be modeled with an energy-momentum tensor corresponding to finite-width topological defects~\cite{Gregory:1995hd} or magnetic flux tubes~\cite{Dowker:1993bt}.

The conserved energy for an accelerating black hole has been studied in an anti-de Sitter (AdS) background~\cite{Appels:2016uha, Appels:2017xoe, Anabalon:2018qfv, Anabalon:2018ydc}.
This approach is motivated by the holographic principle, which enables the derivation of conserved energy from the AdS boundary. Hence, an AdS boundary is essential for defining conserved energy in these scenarios. However, in the case of an accelerating Schwarzschild black hole, obtaining the conserved energy using the holographic principle proves challenging due to its asymptotically flatness.

Assuming the first law of thermodynamics $\it{ ab~initio}$, the conserved energy for the C-metric could be derived through the integration of $T \dd S$~\cite{Astorino:2016xiy}. In Ref.~\cite{Ball:2020vzo}, the local form of the first law of thermodynamics was proven using the covariant phase space formalism~\cite{Wald:1993nt,Iyer:1994ys}, and the conserved energy of the accelerating black hole was chosen to be consistent with the Christodoulou–Ruffini formula~\cite{Christodoulou:1971pcn}.
In fact, a primary difficulty arises in defining the conserved energy of accelerating black holes from the covariant phase space method, which is complicated by the presence of the acceleration horizon.

To circumvent the complexities linked to the exotic structure of the accelerating black hole, a quasilocal formalism for conserved energy can be used. Unlike global asymptotic methods, the quasilocal approach offers a strong framework for defining a conserved charge within a finite spacetime region. The conserved energy using the quasilocal method can be derived covariantly by connecting the Abbott-Deser-Tekin (ADT) current~\cite{Abbott:1981ff,Abbott:1982jh,Deser:2002rt,Deser:2002jk} with the linearized Noether current at the off-shell level~\cite{Kim:2013zha}.
In this work, we present a detailed derivation of the conserved energy directly from the quasilocal formalism using the off-shell ADT conserved current, and subsequently demonstrate the thermodynamic first law.

For the metric, it is axisymmetric due to the presence of conical singularities along the azimuthal axis. The energy required for the cloning of information is expected to depend on the polar angle, which is different behavior from that of spherically symmetric black holes. We will investigate the angle-dependence of the required energy for information cloning and study whether black hole complementarity can be well-defined, irrespective of the polar angle. To this end, we derive thermodynamic quantities such as the conserved energy for an accelerating Schwarzschild black hole using the quasilocal formalism~\cite{Kim:2013zha}.
We then explore the validity of BHC for the black hole by performing two thought experiments introduced by Susskind and Thorlacius~\cite{Susskind:1993if},
and Hayden and Preskill~\cite{Hayden:2007cs}.

The organization of the paper is as follows:
In Sec.~\ref{sec:The quasilocal conserved energy of the asymptotically flat accelerating black hole}, we derive the thermodynamic quantities of the accelerating Schwarzschild black hole.
In Sec.~\ref{sec:Black hole complementarity in the accelerating black hole},
we study BHC using the thermodynamic quantities by executing the thought experiment proposed by Susskind and Thorlacius. We also investigate BHC considering the scrambling time according to Hayden-Preskill's criteria. Conclusions and discussions are presented in Sec.~\ref{sec:conclusion}.
Finally, in Appendix~\ref{appendix:The thought experiment with the temperature at the acceleration horizon}, we extend the analysis of BHC by incorporating the effects of thermal radiation from the acceleration horizon.

\section{thermodynamic quantities}
\label{sec:The quasilocal conserved energy of the asymptotically flat accelerating black hole}
In this section, we calculate the energy, entropy, and temperature of the accelerating Schwarzschild black hole.
Let us start with the Einstein-Hilbert action defined as
\begin{equation}
	\label{eq:EH action}
	I = \frac{1}{16\pi G}\int \dd[4]{x} R.
\end{equation}
The vacuum solution to the Einstein equation of motion can be obtained as~\cite{Griffiths:2005qp,Griffiths:2009dfa}
\begin{equation}
	\label{eq:acc metric}
	\dd s^2 = \frac{1}{\left( 1-A r \cos \theta \right)^2}\left[ -f(r)N^2\dd t^2 + \frac{1}{f(r)}\dd r^2 + \frac{r^2}{g(\theta)}\dd \theta^2 + \frac{ g(\theta)r^2 \sin^2 \theta}{K^2} \dd \phi^2 \right],
\end{equation}
where $f(r) = (1-A^2r^2)\left( 1-\frac{2GM}{r} \right)$ and $ g(\theta) = 1 - 2AGM \cos\theta$,
with $M$ and $A$ being the mass and acceleration parameters.
The parameter $N$ is the dimensionless time-scaling factor that depends on the parameters $M$, $A$, and $K$~\cite{Anabalon:2018qfv,Anabalon:2018ydc,Cassani:2021dwa,Kim:2023ncn}.
The event horizon and the acceleration horizon of the black hole are located at $r_h= 2GM$ and $r_A=\frac{1}{A}$, respectively. The conformal infinity is located at $r = \frac{1}{A \cos \theta}$, where $0< \theta < \frac{\pi}{2}$.
To ensure that the event horizon of the black hole lies within the conformal infinity, we assume $ r_h <r_{A}$.
At the spacetime poles $\theta =0$ and $\theta = \pi$, the deficit angles associated with the parameter $K$ are
calculated as
$\Delta \phi\big|_{\theta=0} = 2\pi \left( 1- \frac{1}{K}(1-2A G M) \right)$
and $\Delta \phi\big|_{\theta=\pi} = 2\pi \left( 1- \frac{1}{K}(1+2A G M) \right)$.
The particular choice of $K = 1\pm 2AGM$ eliminates one of the deficit angles;
however, we assume $K$ to be an arbitrary, non-vanishing constant.

By choosing the timelike Killing vector as $\xi = \partial_t$,
we can calculate the black hole temperature at the event horizon as
\begin{equation}
	\label{eq:acc ads temp}
	T_H = \frac{N(1-4A^2G^2M^2)}{8\pi G M},
\end{equation}
and also obtain the entropy of the black hole from Wald's entropy formula~\cite{Wald:1993nt,Wald:1999wa} as
\begin{equation}
	\label{eq:acc ads entropy}
	S = -2\pi \int_{\mathcal{H}} \dd^2x \sqrt{|h|}\pdv{L}{R_{\mu\nu\rho\sigma}}\epsilon_{\mu\nu}\epsilon_{\rho\sigma} = \frac{4\pi GM^2}{K(1-4A^2G^2M^2)}.
\end{equation}
Additionally, the quasilocal energy $E$ is defined by the surface integral, which can be deformed freely without changing the charge, as long as we do not pass through stress-energy sources. If we take the surface as the horizon of the black hole,
Wald's formulation~\cite{Wald:1993nt,Wald:1999wa} gives us the charge as
\begin{equation}
	\label{first law}
	\dd E = T_H \dd S.
\end{equation}
The first law of thermodynamics can always be obtained by considering the horizon as the surface of the quasilocal charge.

Before we calculate the quasilocal energy explicitly, let us consider the integrability condition to determine whether $E$ can be well-defined object~\cite{Astorino:2016xiy,Astorino:2016ybm,Ball:2020vzo}.
In Eq.~\eqref{first law}, the integrability condition is required to be $0 = \dd(T_H\dd S) = \dd T_H \wedge \dd S$,
where we used the fact that $\dd^2 E = 0$.
The general solution to this condition is given as
\begin{equation}
	\label{eq:general normalization}
	N = \frac{2GM}{1-4A^2G^2M^2}H(S),
\end{equation}
where $H(S)$ is an arbitrary differentiable function.
In particular, the function $H(S)$ is chosen as
\begin{equation}
	\label{eq:function H}
	H(S) =\sqrt{\frac{\pi}{G S}}= \frac{\sqrt{K(1-4A^2G^2M^2)}}{2GM}
\end{equation}
so that we have
\begin{equation}
	\label{eq:norm factor}
	N(M, A, K) = \sqrt{\frac{K}{1-4A^2G^2M^2}}.
\end{equation}
Plugging Eq.~\eqref{eq:norm factor} into Eq.~\eqref{eq:acc ads temp}, we obtain
\begin{equation}
	\label{eq:tem}
	T_H = \frac{\sqrt{K(1-4A^2G^2M^2)}}{8\pi GM}
\end{equation}
which recovers the Hawking temperature of the Schwarzschild black hole when $K \to 1$ and $A\to0 $. In particular, Eq.~\eqref{eq:tem} for $K = 1+2AGM$ matches the temperature expression in Ref.~\cite{Astorino:2016xiy} in the absence of angular momentum and electric charge.
The temperature expression \eqref{eq:tem} is somewhat general.

Now, the quasilocal charge corresponding to the Killing vector $\xi$, linearized with respect to an arbitrary background, is given by~\cite{Kim:2013zha}
\begin{equation}
	\label{eq:ADT charge general}
	Q_{\rm ADT}(g \,;\,\xi,\delta g) = \frac{1}{16\pi G}\int_{B}\dd[2]x_{\mu\nu} K^{\mu\nu}_{\rm ADT}= \frac{1}{16\pi G}\int_{B}\dd[2]x_{\mu\nu}\left( \delta K^{\mu\nu}(g\,;\,\xi) - 2\xi^{[\mu}\Theta^{\nu]}(g\,;\,\delta g) \right),
\end{equation}
where $\dd[2]{x_{\mu\nu}} = \frac{1}{4} \epsilon_{\mu\nu \alpha\beta}\dd x^\alpha \wedge \dd x^\beta $ with $\epsilon_{tr\theta\phi}=-1$.
On the Cauchy surface defined as
$\Sigma = \{(r ,\theta,\phi)\mid 2GM <r < \rho,\ 0<\theta<\pi,\ 0<\phi<2\pi\}$,
we require $\rho < \frac{1}{A}$ so that the Cauchy surface $\Sigma$ does not extend to the conformal infinity at $\theta=0$.
The partial boundary $B$ of the Cauchy surface $\Sigma$ is defined by
\begin{align}
	\label{eq:cauchy bdy}
	B &= \partial \Sigma_0 \cup \partial \Sigma_\rho \cup \partial \Sigma_\pi,\\
	\partial \Sigma_\rho &= \{(\rho,\theta,\phi)\mid \ 0<\theta<\pi,\ 0<\phi<2\pi\},\\
	\partial \Sigma_0 &= \{(r,0,\phi)\mid \ 2GM<r<\rho,\ 0<\phi<2\pi\},\label{eq:0 cauchy bdy}\\
	\partial \Sigma_\pi &= \{(r,\pi,\phi)\mid \ 2GM<r<\rho,\ 0<\phi<2\pi\}.\label{eq:pi cauchy bdy}
\end{align}
In Eq.~\eqref{eq:ADT charge general}, $K^{\mu\nu}$ is the off-shell Noether potential and $\Theta^\mu$ is the surface term generated from the metric variation of the action \eqref{eq:EH action}:
$K^{\mu\nu}(g\,;\,\xi) = 2\sqrt{-g}\nabla^{[\mu}\xi^{\nu]}$ and
	$\Theta^\mu(g\,;\,\delta g) = 2\sqrt{-g}\left( g^{\mu[\lambda}g^{\kappa]\nu}\nabla_{\kappa}\delta g_{\nu\lambda} \right)$,
where the metric variation is arbitrary at the off-shell level.
After performing the variation, the metric is varied along a one-parameter path in the solution space in terms of $\lambda$ by replacing $M$ with $ \lambda M$, where $0\leq\lambda\leq1$.
Then, the off-shell ADT potential $K^{\mu\nu}_{\rm ADT} =  \delta_\lambda K^{\mu\nu}(g\,;\,\xi) - 2\xi^{[\mu}\Theta^{\nu]}(g\,;\,\delta_\lambda g)$ is calculated as
\begin{align}
	K^{tr}_{\rm ADT} &=-\frac{4GM(1-A^2r^2)\sin\theta}{(1-Ar\cos\theta)^3\sqrt{K(1-4A^2G^2M^2\lambda^2)}},\label{eq:ADT potential tr}\\
	K^{t\theta}_{\rm ADT} &= -\frac{AGM(2+6\cos 2\theta - Ar(7\cos \theta + \cos 3\theta))}{2(1-Ar\cos\theta)^3\sqrt{K(1-4A^2G^2M^2\lambda^2)}}.\label{eq:ADT potential ttheta}
\end{align}
Inserting Eqs.~\eqref{eq:ADT potential tr} and \eqref{eq:ADT potential ttheta} into Eq.~\eqref{eq:ADT charge general}, one can obtain
\begin{align}
	Q_{\rm ADT}(g;\lambda)
	&= \frac{1}{16\pi G}\left( \int_{\partial \Sigma_{0,\lambda}} + \int_{\partial \Sigma_\rho}+ \int_{\partial \Sigma_{\pi,\lambda}} \right)\dd[2]x_{\mu\nu}K^{\mu\nu}_{\rm ADT}\nonumber\\
	&= \frac{M}{(1-4A^2G^2M^2\lambda^2)^{3/2}\sqrt{K}},\label{eq:acc ads adt charge}
\end{align}
where the conical singularities on $\partial \Sigma_{0,\lambda}$ and $\partial \Sigma_{\pi,\lambda}$ are treated as boundaries:
$\partial \Sigma_{0,\lambda} = \{(r,0,\phi)\mid \ 2GM\lambda<r<\rho,\ 0<\phi<2\pi\}$ and
	$\partial \Sigma_{\pi,\lambda} = \{(r,\pi,\phi)\mid \ 2GM\lambda<r<\rho,\ 0<\phi<2\pi\}$.
In Eq.~\eqref{eq:acc ads adt charge}, the angular component of the surface term $\Theta^{\theta}$ in
the first and third integrations does not vanish along the azimuthal axis at $\theta = 0$ and $\theta=\pi$, so the integration along these axes must be properly accounted for.

Ultimately, the quasilocal energy for the accelerating Schwarzschild black hole is given by
\begin{equation}
	\label{eq:acc energy}
	E = \int_{0}^{1}Q_{\rm ADT}(g;\lambda)\dd \lambda = \frac{r_Ar_h}{2G\sqrt{K(r_A^2 - r_h^2)}},
\end{equation}
which is independent of the radial direction.
In Eq.~\eqref{eq:acc energy}, the energy expression is proportional to $\frac{1}{\sqrt{K}}$ whereas the entropy \eqref{eq:acc ads entropy} is proportional to $\frac{1}{K}$ due to the azimuthal integration.
However, in contrast to the entropy, the conserved energy \eqref{eq:acc energy} is affected by the choice of $N$ through $\sqrt{-g}$ in the Noether potential $K^{\mu\nu}$ and the surface term $\Theta^\mu$.
From the integrability condition, the factor \eqref{eq:norm factor} turns out to be proportional to $\sqrt{K}$.
Consequently, performing the azimuthal integration along with the factor $N$ yields $\frac{1}{K}\times \sqrt{K} = \frac{1}{\sqrt{K}}$ in Eq.~\eqref{eq:acc energy}.
In particular, this energy expression \eqref{eq:acc energy} reduces to the mass of the Schwarzschild black hole
when $K \to 1$ and $A\to0 $.
From now on, we will set $K=1$ for convenience.

\section{Black hole complementarity}
\label{sec:Black hole complementarity in the accelerating black hole}
In this section, we consider BHC for the accelerating Schwarzschild black hole using thermodynamic quantities.
First, we define the Kruskal coordinates as $x^{\pm} = \pm \frac{1}{\kappa}e^{\pm \kappa\sigma^{\pm}} = \pm \frac{1}{\kappa}e^{\pm \kappa(t\pm r^\ast)}$, where
$r^\ast = \int \frac{1}{N f(r)}\dd r = \frac{1}{2\kappa_+}\log(1+Ar) - \frac{1}{2\kappa_-}\log(1-Ar) + \frac{1}{2\kappa} \log(\frac{r}{2GM} - 1)$,
with $\kappa_\pm = A\sqrt{\frac{1\pm 2AGM}{1\mp 2AGM}}$ and $\kappa = \frac{\sqrt{1-4A^2G^2M^2}}{4GM}$.
The metric \eqref{eq:acc metric} with fixed angles $\theta$ and $\phi$ can be rewritten as
\begin{equation}
	\label{eq:metric Kruskal}
	\dd s^2 = -\frac{2GMN^2(1-A^2r^2)^{3/2}}{(1-Ar\cos \theta)^2r} \exp\left[ -\frac{1}{4AGM}\log(\frac{1+Ar}{1-Ar}) \right]\dd x^+\dd x^-.
\end{equation}
In the Kruskal coordinates, the curvature singularity of the black hole is located at $x^+x^- = 16G^2E^2$.

Note that thermal radiation is emitted from both the event horizon and the acceleration horizon. However, we focus exclusively on radiation from the event horizon, since Alice will cross into it and thus her information will be located inside that horizon rather than the accelerating horizon. The contribution of thermal radiation from the acceleration horizon will be discussed in Appendix~\ref{appendix:The thought experiment with the temperature at the acceleration horizon}.
\begin{center}
	\begin{figure}
		\begin{tikzpicture}
			\def\a{2.5}
			\def\x{.4}
			\def\z{.1}
			\def\y{.6}
			\draw[thick,->] (-.5,-.5) -- (4,4) node[above right] {$x^+$};
			\draw[thick,->] (.5,-.5) -- (-4,4) node[above left] {$x^-$};
			\draw[decorate,decoration={zigzag,segment length=2mm,amplitude=.5mm}] plot[variable=\y,domain={\a/(4*sqrt(2))}:{(4*sqrt(2))}] ({(\y-\a/\y)/sqrt(2)},{(\y+\a/\y)/sqrt(2)});
			\draw[-latex,very thick] ({1-.18},.18) to[out=135,in=-90] node[below,sloped] {\scriptsize Alice} (0,2.2);
			\draw[dotted,thick] (0.2,1.1) -- ({.2+\y/(sqrt(2))},{1.1-\y/(sqrt(2))});
			\draw[<-] (.6,.6) to[out=-45,in=90] (1.5,-.4) node[below] {\scriptsize $\Delta x^+_{\rm A}$};
			\draw[-latex,very thick] ({3+.32},{3-.32}) -- ({(3*sqrt(2)-\a/(3*sqrt(2)))/sqrt(2)},{(3*sqrt(2)+\a/(3*sqrt(2)))/sqrt(2)})node[midway,sloped,below] {\scriptsize Bob} ;
			\draw[->] ({(1+\x - \a/3)/2},{(1+\x + \a/3)/2}) --  ({(3*sqrt(2)-\a/(3*sqrt(2)))/sqrt(2)},{(3*sqrt(2)+\a/(3*sqrt(2)))/sqrt(2)}) node[midway,sloped,below=1mm] {\scriptsize $\Delta \epsilon$};
			\draw[decorate,decoration={brace,mirror,amplitude=3pt}] ({(1+\z)/2 + 1/(2*sqrt(2))},{(1+\z)/2 - 1/(2*sqrt(2))}) -- ({3.1 -\z + 1/(2*sqrt(2))},{3.1-\z - 1/(2*sqrt(2))}) node[midway,rotate=45,below=1mm] {\scriptsize $\Delta t$};
			\draw (0,2.5)node[above] {\scriptsize $x^+x^- = 16G^2E^2$};
			\draw[decorate,decoration={snake, segment length=3mm, amplitude=.5mm},->] ({(1+10*\z)/2 + 1/(2*sqrt(2))-.12},{(1+10*\z)/2 - 1/(2*sqrt(2))+.12}) -- ({3.5 -10*\z + 1/(2*sqrt(2))-.12},{3.5-10*\z - 1/(2*sqrt(2))+.12});
		\end{tikzpicture}
		\caption{Alice falls into the black hole with her information at $x^+_{\rm A}$ where $\Delta x^+_{\rm A}$ that depends on Alice’s initial conditions,
and Bob retrieves the information via Hawking radiation outside the horizon.
			Once the information is reconstructed during $\Delta t$ which can be the Page time $t_{\rm P}$ or the scrambling time $t_{\rm scr}$
depending on experiments, Bob jumps into the black hole towards the singularity, represented by the zigzag line.
			Alice sends her information with the frequency $\Delta \epsilon$ to Bob before he reaches the singularity.}
		\label{fig:acc}
	\end{figure}
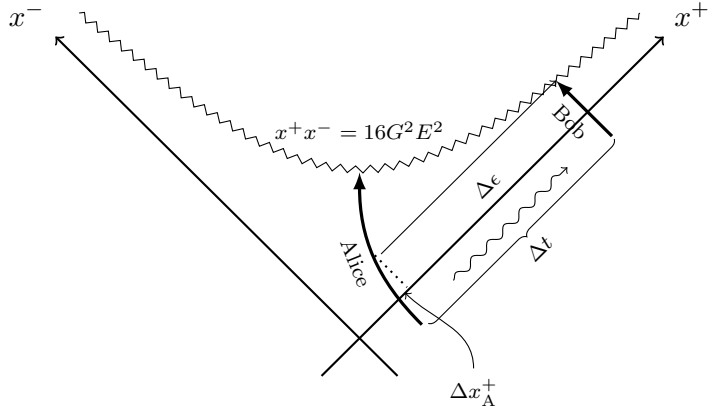
\end{center}
\subsection{The thought experiment based on the Page time}
\label{sec:The thought experiment based on the Page time}
Let us begin with the thought experiment based on the Page time~\cite{Susskind:1993mu}.
Alice falls freely into the event horizon of a collapsing black hole, carrying with her information.
Meanwhile, Bob hovers just outside the black hole, gathering the information emitted in the form of early Hawking radiation.
According to Page's calculation~\cite{Page:1993wv}, Bob can recover information about the collapsing matter, including Alice's information, after the Page time has passed.
After obtaining this information from the Hawking radiation, Bob decides to cross the event horizon and approach the curvature singularity. As Bob is falling, Alice sends a message containing her information to Bob before he reaches the singularity, as depicted in Fig.~\ref{fig:acc}.
If Bob receives Alice's message before he encounters the singularity, he would then possess two copies of the same information: one from the Hawking radiation and one directly from Alice. This duplication would result in a violation of quantum unitarity, suggesting an inconsistency in the quantum mechanical description.
However, this paradox may be resolved if the energy required for Alice to send her message to Bob is extremely high, at a super-Planckian scale.
As a result, the super-Planckian energy would prevent Alice from transmitting the message, thereby avoiding the possibility of information duplication and preserving the unitarity of quantum mechanics~\cite{Susskind:1993mu}.

Explicitly, the Page time for an evaporating black hole can be obtained by using the Stefan-Boltzmann law:
\begin{equation}
	\label{eq:SB law}
	\dv{E}{t} = -\mathcal{A}\sigma T_H^4,
\end{equation}
where $\sigma$ is the Stefan-Boltzmann constant and $\mathcal{A}$ is the area of the black hole, identified as $\mathcal{A} = 4G S$.
Using Eq.~\eqref{eq:SB law} along with Eqs.~\eqref{eq:tem} and \eqref{eq:acc energy} for $K=1$, one can calculate the Page time~\cite{Page:1993wv}
\begin{equation}
	\label{eq:Page time}
	t_{\rm P} = -\frac{16\pi^{3/2} \sqrt{G}}{\sigma}\int_{\frac{1}{2}S}^{S} \sqrt{S} \dd S = 16\gamma G^2 E^3,
\end{equation}
where $\gamma = \frac{4\pi^3}{3\sigma}(4-\sqrt{2})$.
After the Page time $\Delta t=t_{\rm P}$ in Fig.~\ref{fig:acc}, Bob enters the black hole at $x^{+}_{\rm B} = \frac{1}{\kappa} e^{\kappa (\sigma^+_{\rm A} + t_{\rm P})} = 4GE e^{4\gamma G E^2}$.
Here, we set Alice's initial location $\sigma^+_{\rm A}$ to zero for convenience.
Since the curvature singularity is located at $x^+x^- = 16G^2E^2$, we can determine the coordinate
when Bob reaches the singularity as $x^{-}_{\rm B} = 4GE e^{-4\gamma G E^2}$.
Hence, if Alice falls freely, the proper time she experiences between the horizon at $x^- = 0$ and $x^- = x^-_{\rm B}$ is calculated
from the metric~\eqref{eq:metric Kruskal} in the near-horizon limit as
\begin{equation}
	\label{eq:Alice proper time in Page}
	\Delta \tau^2 = \frac{2r_A^2 r_h\Delta x_{\rm A}^+}{(r_A - r_h\cos \theta)^2}\left( \frac{r_A - r_h}{r_A+r_h} \right)^{\frac{r_A}{2r_h}}\exp\left[ - \frac{\gamma r_A^2r_h^2}{G(r_A^2 - r_h^2)} \right].
\end{equation}
Using the uncertainty principle $\Delta \tau \Delta \epsilon > \frac{1}{2}$~\cite{Susskind:1993mu}, we obtain
\begin{equation}
	\label{eq:Energy Page}
	\Delta \epsilon^2 > \frac{(r_A - r_h\cos \theta)^2}{8r_A^2 r_h\Delta x_{\rm A}^+}\left( \frac{r_A + r_h}{r_A - r_h} \right)^{\frac{r_A}{2r_h}}\exp\left[\frac{\gamma r_A^2r_h^2}{G(r_A^2 - r_h^2)} \right],
\end{equation}
where $\Delta x_{\rm A}^+ $ is a finite numerical constant that depends on Alice's initial data~\cite{Hayden:2007cs}.
For $\ell_{\rm P} \ll r_h <r_A$, we assume that both $r_h$ and $r_A$ are finite, and $r_A - r_h $ is also finite. Then, the energy scale in Eq.~\eqref{eq:Energy Page} becomes manifestly of the super-Planckian order, independent of the polar angle:  $\Delta \epsilon^2 >\mathcal{O}\left( \exp\left[\frac{\gamma r_A^2r_h^2}{4\ell_{\rm P}^2(r_A^2 - r_h^2)} \right] \right)$.
If we take the Schwarzschild limit $r_A \to \infty $ with the finite $r_h$, the inequality \eqref{eq:Energy Page} reduces to that of a Schwarzschild black hole: $\Delta \epsilon^2 > \frac{1}{8r_h\Delta x_{\rm A}^+}\exp(\gamma\frac{ r_h^2}{G}) = \mathcal{O}\left( \exp(\gamma\frac{ r_h^2}{\ell_{\rm P}^2}) \right)$, which is also super-Planckian.
Therefore, Alice cannot transmit her information to Bob, ensuring that information cloning is avoided.

\subsection{The thought experiment based on the scrambling time}
\label{sec:The thought experiment based on the scrambling time}
Let us now consider the scenario in which the black hole is already maximally entangled with a quantum memory held by Bob~\cite{Hayden:2007cs}. In this situation, Bob is capable of decoding the initial quantum state encoded within the black hole.
Alice, carrying her own information that is still unknown to Bob, freely falls into the event horizon. While Bob remains outside, he continues to gather the information emitted through Hawking radiation. According to the proposal by Hayden and Preskill~\cite{Hayden:2007cs}, Bob can recover Alice's information after the scrambling time has passed.
Once Bob has retrieved the information, he crosses the event horizon and heads toward the singularity. Meanwhile, Alice attempts to transmit a message containing her information to Bob before he reaches the singularity.

The scrambling time is typically given by \cite{Hayden:2007cs,Sekino:2008he}
\begin{equation}
	\label{eq:Scramb time}
	t_{\rm scr} \sim \frac{1}{2\pi T_H}\log S = 4GE \log(4\pi G E^2),
\end{equation}
where we used Eqs.~\eqref{eq:tem} and \eqref{eq:acc energy}.
After the scrambling time $t_{\rm scr}$ in Fig.~\ref{fig:acc}, Bob jumps into the horizon, and his coordinate is given by
$x^{+}_{\rm B} = \frac{1}{\kappa} e^{\kappa (\sigma^+_{\rm A} + t_{\rm scr})}= 16\pi G^{2} E^{3}$,
where $\sigma_{\rm A}^+$ is set to zero for convenience.
Since the curvature singularity is located at $x^+x^- = 16G^2E^2$, we can find that the coordinate
for Bob when he reaches the singularity is $x^{-}_{\rm B} = \pi^{-1}E^{-1}$.
Then, the proper time $\Delta \tau$ from the horizon
at $x^{-} = 0$ to $x^- = x^-_{\rm B}$ can be calculated from the metric~\eqref{eq:metric Kruskal} near the horizon as
\begin{equation}
	\label{eq:Alice proper time in scr}
	\Delta \tau^2 = \frac{2G(r_A^2-r_h^2)\Delta x_{\rm A}^+}{\pi r_h(r_A - r_h\cos \theta)^2}\left( \frac{r_A - r_h}{r_A+r_h} \right)^{\frac{r_A}{2r_h}}.
\end{equation}
Using the uncertainty principle~\cite{Susskind:1993mu}, we obtain
\begin{equation}
	\label{eq:Energy n times scramb}
	\Delta \epsilon^2 > \frac{\pi r_h(r_A - r_h\cos \theta)^2}{8G(r_A^2-r_h^2)\Delta x_{\rm A}^+}\left( \frac{r_A + r_h}{r_A - r_h} \right)^{\frac{r_A}{2r_h}}.
\end{equation}
We assume that both $r_h$ and $r_A$ are finite, with $r_h \gg  \ell_{\rm P}$ semiclassically, and that $r_A - r_h $ is also finite to avoid an extremal case.
Under these conditions, the factors $\left( \frac{r_A + r_h}{r_A - r_h} \right)^{\frac{r_A}{2r_h}}$, $ (r_A - r_h\cos \theta)$, and $(r_A^2-r_h^2)$ in Eq.~(26) remain finite.
Additionally, $\Delta x_{\rm A}^+$ is a finite constant depending on Alice’s initial conditions.
Therefore, we have $\Delta \epsilon^2 > \text{finite} \times \frac{1}{G} \sim \frac{1}{\ell_{\rm P}^2}= m_{\rm P}^2.$
Hence, the expression \eqref{eq:Energy n times scramb} is of the order $\mathcal{O}\left( \frac{1}{\ell_{\rm P}^2} \right)$, which indeed confirms that it is super-Planckian independent of the polar angle.
This ensures that information cloning is still evaded.
Note that if we take the Schwarzschild limit $r_A \to \infty $ for the
finite $r_h$, the inequality \eqref{eq:Energy n times scramb} reduces to the result of a Schwarzschild black hole: $\Delta \epsilon^2 > \frac{e\pi r_h}{ 8G\Delta x^{+}_{\rm A}} = \mathcal{O}\left( \frac{1}{\ell_{\rm P}^2} \right)$
which is also super-Planckian.
Consequently, Alice cannot transmit her information to Bob, ensuring that
duplication of information is avoided.

\section{Conclusion and discussion}
\label{sec:conclusion}
Using the quasilocal formalism for conserved charges, we derived the expression for the energy of the accelerating Schwarzschild black hole, which reduces to the mass of the Schwarzschild black hole when $A \to 0$. Utilizing thermodynamic quantities, we determined both the Page time and the scrambling time, and subsequently investigated BHC for the black hole by performing thought experiments based on these two different times. Consequently, the duplication of quantum states is prevented because the energy required to transmit the message exceeds the super-Planckian scale, thereby confirming that black hole complementarity remains valid for the accelerating Schwarzschild black hole.

In our thought experiment, Alice falls into the event horizon rather than the acceleration horizon.
Thus, her quantum information is scrambled only with the degrees of freedom behind the event horizon, thereby Bob can recover Alice's information from the thermal radiation emitted from the event horizon solely.
In other words, the radiation from the acceleration horizon does not carry Alice's information in our setup and therefore does not contribute to Bob's retrieval process.

For the Killing vector, we chose it as $\xi=\partial_t$ without the normalization factor.
It is worth noting that adding a normalization factor for the Killing vector is unnecessary in our calculations.
This is because the metric itself includes the factor $N$, effectively serving as a normalization factor.
If we rescale the time coordinate in Eq.~\eqref{eq:acc metric} and remove the factor $N$ from the metric, the normalization factor $N$ would then appear in the Killing vector as $\xi = N \partial_t$.
In either case, the resulting conserved energy remains unchanged.

Thus far, we have focused on the non-extremal case of the accelerating Schwarzschild black hole. One may wonder what happens if the acceleration horizon approaches the event horizon, \textit{i.e.},  $ \ell_P \ll r_h \sim r_A$.
In Eqs.~\eqref{eq:Energy Page} and~\eqref{eq:Energy n times scramb}, if we introduce a cutoff such that $r_A - r_h = \ell_{\rm P}$, the energy uncertainty becomes $\Delta \epsilon^2 > \mathcal{O}\left( \ell_{\rm P}^{3/2} \exp(\frac{\gamma r_h^3}{2\ell_{\rm P}^3}) \right)$ and $\Delta \epsilon^2 > \mathcal{O}\left( \frac{1}{\ell_{\rm P}^{3/2}} \right)$, respectively.
This indicates that the required energy for the duplication of information remains at the super-Planckian scale, even in the near-extremal limit with a very small temperature $T_H = \mathcal{O}(\sqrt{\ell_{\rm P}})$.
Thus, the current analysis shows that black hole complementarity holds for $ \ell_P \ll r_h \lesssim r_A$.

\appendix
\section{BHC with thermal contributions from the acceleration horizon}
\label{appendix:The thought experiment with the temperature at the acceleration horizon}
One may naturally wonder how the presence of this thermal radiation affects BHC.
To address this, let us define the temperature associated with the thermal radiation emitted from the acceleration horizon as
\begin{equation}
	\label{tem2}
	T_A = \frac{\kappa_-}{2\pi} = \frac{A}{2\pi}\sqrt{\frac{1-2AGM}{1+2AGM}}.
\end{equation}
Comparing Eq.~\eqref{tem2} to the Hawking temperature in Eq.~\eqref{eq:tem}, we find $\frac{T_H}{T_A} = \frac{1+2AGM}{4AGM} > 1$ since $2AGM < 1$.
This inequality shows that $T_H$ is greater than $T_A$, implying that the net thermal flux is directed from the event horizon towards the acceleration horizon.
As a consequence, the black hole will continue to evaporate under this net flux.

As shown in Refs.~\cite{Shankaranarayanan:2003ya,Urano:2009xn}, spacetimes with multiple horizons can produce a combined thermal spectrum described by the inverse harmonic sum of the individual horizon temperatures.
This result follows from analyzing the ratio of emitted to absorbed particle fluxes.
To incorporate the contribution of the thermal radiation emitted from both the event horizon and the acceleration horizon, we assume an effective temperature analogous to that of the Schwarzschild–de Sitter black hole as
\begin{equation}
	\label{eq:eff temp}
	T_{\rm eff} = \frac{T_H T_A}{T_H + T_A} = \frac{A\sqrt{1-4A^2G^2M^2}}{2\pi (1+6AGM)}.
\end{equation}

Next, we assume that in the region between the two horizons, the relevant temperature is given by the effective temperature $T_{\rm eff}$.
The Stefan–Boltzmann law is then modified accordingly
\begin{equation}
	\label{}
	\dv{E}{t} = -\mathcal{A} \sigma T_{\rm eff}^4.
\end{equation}
Repeating the calculation similar to Eq.~\eqref{eq:Page time}, we obtain the effective Page time as
$t_{\rm P} \approx \frac{\widetilde{\gamma}}{A^4 G^2 M} + \mathcal{O}\left( \frac{1}{G} \right)$,
where $\widetilde{\gamma} = \frac{2\pi^3}{\sigma(\sqrt{2}+1)}$.
The required energy for Alice to send her information to Bob is then found to be
\begin{equation}
	\Delta \epsilon^2 > \mathcal{O}\left(\exp\left[\frac{\widetilde{\gamma} r_A^3\sqrt{r_A^2 - r_h^2}}{r_h^2 \ell_{\rm P}^2}\right]\right) \gg  \mathcal{O}\left( \frac{1}{\ell_{\rm P}^2} \right).
\end{equation}
This shows that duplicating Alice’s quantum information requires an energy that exceeds super-Planckian, thereby evading any violation of the no-cloning principle in quantum mechanics.

We now define the effective scrambling time as
\begin{equation}
	\label{}
	t_{\rm scr} = \frac{1}{2\pi T_{\rm eff}}\log S = \frac{r_A^2 + 3r_Ar_h}{\sqrt{r_A^2 - r_h^2}}\log(4\pi G E^2).
\end{equation}
A similar calculation shows that the required energy for Alice to send her information to Bob can be expressed as
\begin{equation}
	\Delta \epsilon^2 > \mathcal{O}\left(\left(\frac{1}{\ell_{\rm P}}\right)^{\frac{r_A}{r_h} + 3}\right) >  \mathcal{O}\left( \frac{1}{\ell_{\rm P}^2} \right).
\end{equation}
This again demonstrates that the energy scale needed is manifestly super-Planckian, independent of the polar angle.
Consequently, the duplication of Alice's quantum information is prevented, and the qualitative nature of BHC
remains unchanged.
\acknowledgments
This work was supported by the National Research Foundation of Korea(NRF) grant funded by the Korea government(MSIT) (NRF-2022R1A2C1002894) and
by Basic Science Research Program through the National Research Foundation of Korea(NRF) funded by the Ministry of Education through the Center for Quantum Spacetime (CQUeST) of Sogang University (NRF-2020R1A6A1A03047877).


\bibliographystyle{JHEP}       

\bibliography{reference}

\end{document}